\documentclass[
aps,%
12pt,%
final,%
notitlepage,%
oneside,%
onecolumn,%
nobibnotes,%
nofootinbib,
superscriptaddress,%
noshowpacs,%
centertags]%
{revtex4}

\usepackage{amssymb,latexsym,amsmath,epsfig}
\newcommand{\eq}[1]{\begin{align} #1 \end{align}}

\begin{document}
%
\title{Multiplicity fluctuations in relativistic gases. \\
       From simple models to experiment.}
\author{\firstname{Viktor} \surname{Begun}}
\affiliation{Bogolyubov Institute for Theoretical Physics, Kiev,
Ukraine.}
%
\noaffiliation
\begin{abstract}
The aim of this paper is to give a short overview for the set of
publications considering recently found effect of non-equivalence
of multiplicity fluctuations in relativistic gases with globally
conserved charge and energy.

\end{abstract}
\maketitle

\section{Introduction}
It was suggested to use the statistical approach to strong
interactions more than 50 years ago \cite{Fermi}, \cite{Landau},
\cite{Hagedorn}. It appeared to be surprisingly successful in
describing experimental results on hadron production properties in
nuclear collisions at high energies (see  e.g. Ref.~\cite{PBM},
\cite{param}, \cite{EbyE} and references therein). This motivates
a rapid development of statistical models and it raises new
questions, previously not addressed in statistical physics. In
particular, an applicability of the models formulated within
various statistical ensembles. Recently, it was found that global
conservation laws suppress multiplicity fluctuations and this
suppression survive even in thermodynamic limit \cite{ce1},
\cite{mce1}. This unexpected result gave rise to the set of
publications on this subject \cite{ce1}-\cite{saddle_point}. This
paper gives a short overview starting from simple models
\cite{ce1}, \cite{mce1} to the recently found experimental
confirmation of this effect \cite{Lungwitz}, \cite{CE-HSD}.

\section{Multiplicity fluctuations}
\begin{figure}[h!]
\begin{center}
 \includegraphics[width=0.9\textwidth]{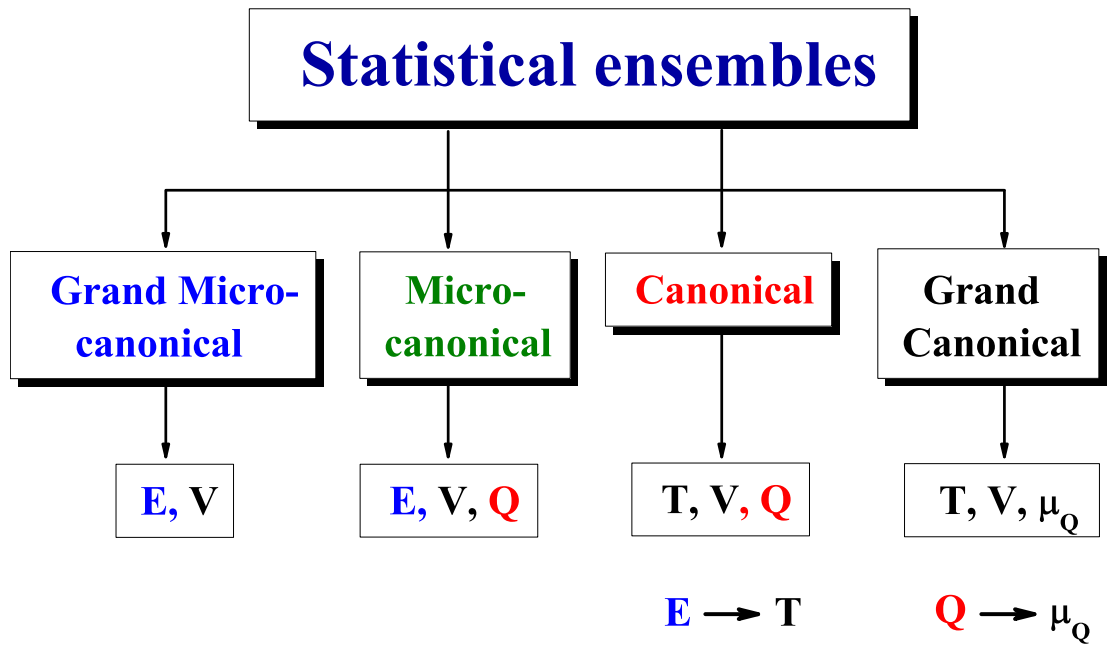}
 \vspace{-1cm}
 \caption{Conservation laws in different statistical
ensembles.}\label{fig-ens}
\end{center}
\end{figure}

Multiplicity fluctuations can be quantified by the scaled
variance. For positively, and negatively, charged particles the
scaled variance reads:
 \eq{\label{W_def}
 \omega^{\pm}
 \equiv \frac{\langle N_{\pm}^2 \rangle - \langle N_{\pm} \rangle^2}{\langle N_{\pm}
 \rangle} \;,
 }
where angular brackets $\langle\;\;\rangle$ means averaging. The
scaled variance is a useful measure, because for Poisson
distribution it equals 1, independently of its mean value:
 \eq{
 \omega_{poisson}^{\pm}= 1
 }
Thus, the scaled variance says how much the studied system is
different from Poisson distribution. Experimentally, the averaging
in the Eq.~(\ref{W_def}) means the averaging on event-by-event
basis: a given observable is measured in each collision event and
the fluctuations are evaluated for the selected set of these
events (see, e.g., review \cite{EbyE}). To calculate a statistical
"background" for multiplicity fluctuations one has to choose a
statistical ensemble for this calculation: grand canonical (GCE),
canonical (CE), microcanonical (MCE) or grand microcanonical
(GMCE), see Fig.~\ref{fig-ens}.
Usually authors do not make the difference between MCE and GMCE
and call both as microcanonical ensemble. We introduce different
names following the suggestion of a referee, because we analyze
and compare different ensembles in details.

The choice of an ensemble depends on the experimental situation.
If one exactly knows the energy, volume and charge of the system
then such a system should be described in the MCE. Sometimes
temperature of a system with exactly known electric charge can be
measured much easier then its whole energy. Then such a system
should be treated in CE, etc... In practice, calculations in CE
and especially in GMCE and MCE are very difficult thus real
calculations are always performed in GCE. One usually refers here
to the textbook statement that all ensembles are equivalent in
thermodynamic limit.

This is the case for particle multiplicities. Different ensembles
are equivalent if one choose a temperature and chemical potentials
in a way that some exactly fixed variable in one ensemble equals
to its adjoint average value in another ensemble, e.g. temperature
$T$ is defined from the condition $E_{m.c.e.}=\langle
E\rangle_{c.e.}$, and chemical potential $\mu_Q$ from the
condition $Q_{c.e.}=\langle Q\rangle_{g.c.e.}$, etc..., see
Fig.~\ref{fig-ens}. However the equivalence of statistical
ensembles does not apply to scaled variances. This was firstly
found in \cite{ce1} and will be illustrated below.

\subsection{Canonical ensemble}

As a simplest example, let us consider a relativistic system in
equilibrium which consists of one sort of positively, $N_+$, and
negatively charged particles\footnote{e.g. $\pi^+$ and $\pi^-$
mesons}, $N_-$, with total charge equal to $Q_{c.e.}=N_+-N_-$. In
the case of the Boltzmann ideal gas (the interactions and quantum
statistics effects are neglected) in the volume $V$ and at
temperature $T$ the GCE and CE partition functions read:
 \eq{\label{Zgce}
 Z_{g.c.e.}(T,V,\mu_Q) &\;=
 \sum_{N_+=0}^{\infty}\sum_{N_-=0}^{\infty}
 \frac{(\lambda_+z)^{N_+}}{N_+!}\frac{(\lambda_-z)^{N_-}}{N_-!}
 \; e^{\mu_Q(N_+-N_-)/T}
 \nonumber
 \\
 &\;= \sum_{N_+=0}^{\infty}\sum_{N_-=0}^{\infty} Z_{N_+,N_-}(T,V,\mu_Q)
  \;=\;\exp\left(2z\cosh[\mu_Q/T]\right),
 }
 \eq{
 Z_{c.e.}(T,V,Q)
 &\;=
 \sum_{N_+=0}^{\infty}\sum_{N_-=0}^{\infty}
 \frac{(\lambda_+ z)^{N_+}}{N_+!}\;\frac{(\lambda_- z)^{N_-}}{N_-!}
 \;\delta (Q-[N_+-N_-])
 \nonumber
 \\
 &\;= \sum_{N_+=0}^{\infty}\sum_{N_-=0}^{\infty} Z_{N_+,N_-}(T,V,Q)
 \nonumber
 \\
 &\;= \frac{1}{2\pi}\int_{-\pi}^{+\pi}d\phi\;\;
   \exp\left[\;iQ\phi \;+\;  z\;(\lambda_+\;e^{i\phi}
                   \;+\; \lambda_-\;e^{-i\phi})\;\right]
  \;=\; I_Q(2z),
  \label{Zce}
 }
where $z$ is a single particle partition function:
 \eq{
z= \frac{gV}{2\pi^2}
       \int_{0}^{\infty}p^{2} dp\;
       e^{-\frac{\sqrt{p^{2}+m^{2}}}{T}}
 = \langle N_{\pm}\rangle,
 }
$g$ is a degeneracy factor (number of spin states), $m$ - particle
mass and $\lambda_{\pm}$ are auxiliary parameters that will be set
to unity after calculation of average values. We also labelled the
number of particles in GCE as $\langle N_{\pm}\rangle$. Let us
omit the indexes $c.e.$, $g.c.e.$, etc., for partition function as
the arguments of $Z$ already show to what ensemble it corresponds.
The average values in both the GCE and CE can be calculated as
follows:
 \eq{\label{<N>ce}
 \langle N_{\pm} \rangle
 &\;\equiv\;
 \frac{1}{Z}  \sum_{N_+=0}^{\infty}\sum_{N_-=0}^{\infty} N_{\pm}\; Z_{N_+,N_-}
 \;=\; \left[\frac{1}{Z}\,\lambda_{\pm}\,
 \frac{\partial Z}{\partial \lambda_{\pm}} \right]_{\lambda_{\pm}=1}\;,
 \\
 \label{<N2>ce}
 \langle N_{\pm}^2 \rangle
 &\;\equiv\; \frac{1}{Z}  \sum_{N_+=0}^{\infty}\sum_{N_-=0}^{\infty} N_{\pm}^2\; Z_{N_+,N_-}
 \;=\; \left[\frac{1}{Z}\,\lambda_{\pm}\, \frac{\partial}{\partial \lambda_{\pm}}
  \left(\lambda_{\pm}\, \frac{\partial Z}{\partial \lambda_{\pm}}\right)
 \right]_{\lambda_{\pm}=1} .
  }
In thermodynamic limit, $V\rightarrow\infty$, and for $Q=0$ it
gives:
 \eq{\label{TDlimit-CE}
 \langle N_{\pm} \rangle &=z, &
 \langle N^2_{\pm} \rangle
&=z+z^2,
 \\
 \label{TDlimit-CE2}
\langle N_{\pm} \rangle_{c.e.}
 &\cong z\left(1-\frac{1}{4z}\right),\quad &
 \langle N^2_{\pm}
\rangle_{c.e.} &\cong z^2,
 }
From the definition of the scaled variance (\ref{W_def}) it then
follows \cite{ce1}:
  \eq{\label{Wgce}
  \omega_{g.c.e.}^{\pm} \;\equiv\; \frac{\langle
  N_{\pm}^2 \rangle - \langle N_{\pm} \rangle^2}{\langle N_{\pm} \rangle}
  &\;=\; 1\;,
 \\
 \label{Wce}
 \omega_{c.e.}^{\pm} \;\equiv\; \frac{\langle N_{\pm}^2 \rangle_{c.e.}
 - \langle N_{\pm} \rangle_{c.e.}^2} {\langle N_{\pm} \rangle_{c.e.}}
 &\;=\; \frac{1}{2}\;. }
\begin{figure}[h!]
\begin{center}
 \vspace{-0.5cm}
 \includegraphics[width=0.5\textwidth]{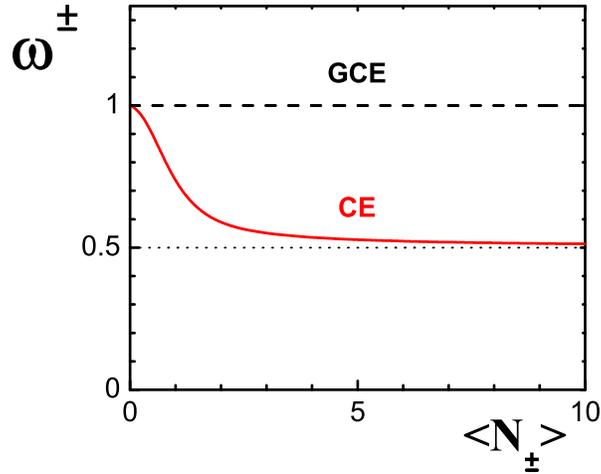}
 \vspace{-0.5cm}
\caption{The scaled variances in GCE (horizontal dashed line) and
in CE (solid line) for $Q=0$ \cite{ce1}.}\label{fig-CE}
\end{center}
\end{figure}

Thus for zero system charge in thermodynamic limit the scaled
variance in CE is two times smaller then in GCE while average
particle numbers are the same, see Eqs.~(\ref{TDlimit-CE}) and
(\ref{TDlimit-CE2}) left, and Eqs.~(\ref{Wgce}), (\ref{Wce}). One
can also see from Fig.~\ref{fig-CE} that the thermodynamic limit
is reached very quickly. The scaled variance $\omega_{c.e.}^{\pm}$
almost reaches its limiting value at $\langle
N_{\pm}\rangle=z\sim5\div 10$.

Multiplicity fluctuations for non-zero charge in multi component
system with two exactly conserved charges, namely electric charge
and baryon number, are considered in \cite{ce2}. The relation
$\omega_{c.e.}=\omega_{g.c.e.}/2$ is preserved in multi component
system if the number of all positively and all negatively charged
particles of different species is the same. Large non-zero charge
$Q>0$ leads to additional suppression of $\omega_{c.e.}^+$ and the
enhancement of $\omega_{c.e.}^-$, while the relation
$\omega_{c.e.}^+<\omega_{c.e.}^-<\omega_{g.c.e.}^{\pm}$ holds.
Additional baryon charge conservation leads to even stronger
suppression of the scaled variance in CE comparing to GCE in
thermodynamic limit.

\subsection{Microcanonical ensemble}

\begin{figure}[h!]
\begin{center}
\includegraphics[width=0.47\textwidth]{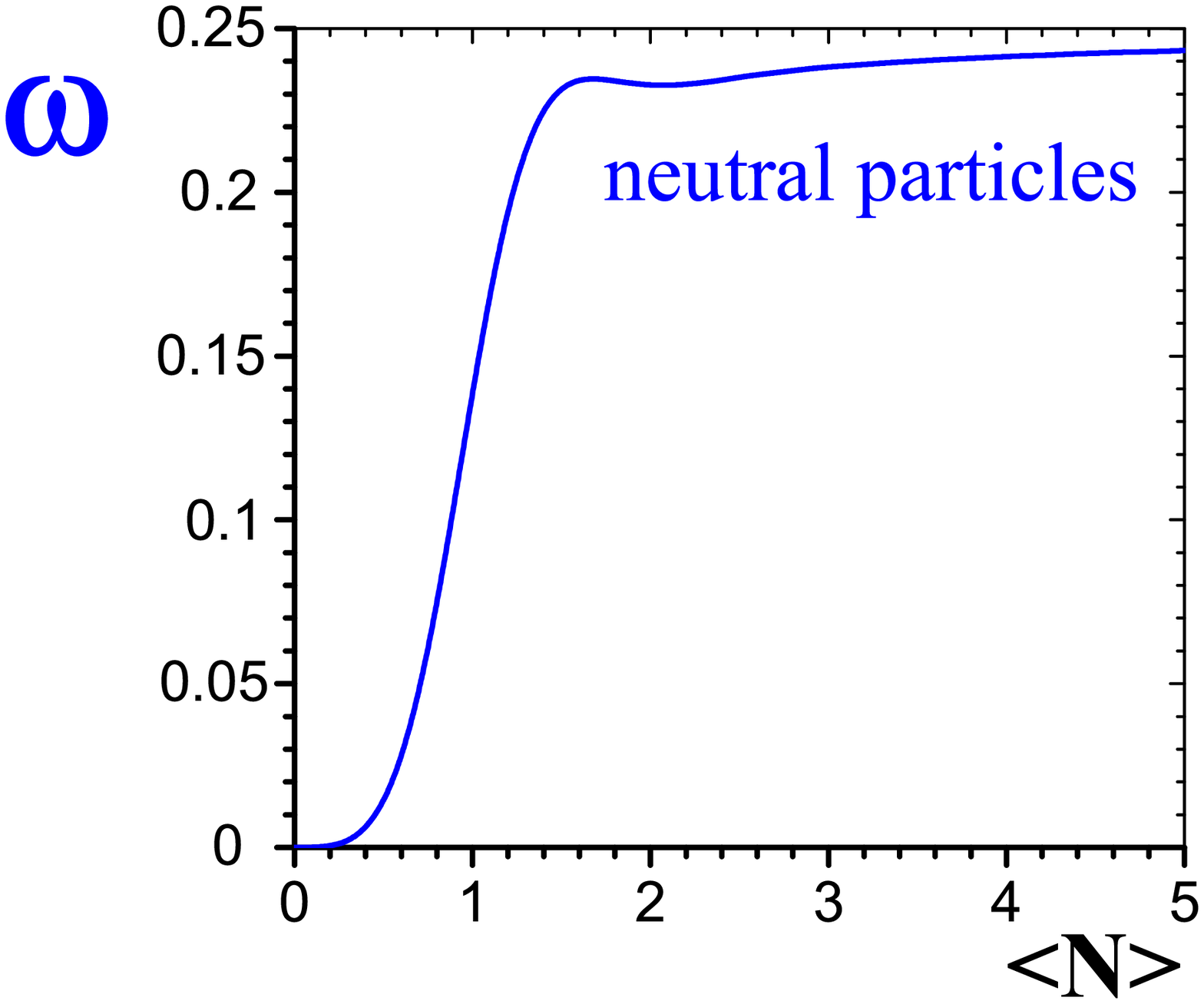}
\includegraphics[width=0.51\textwidth]{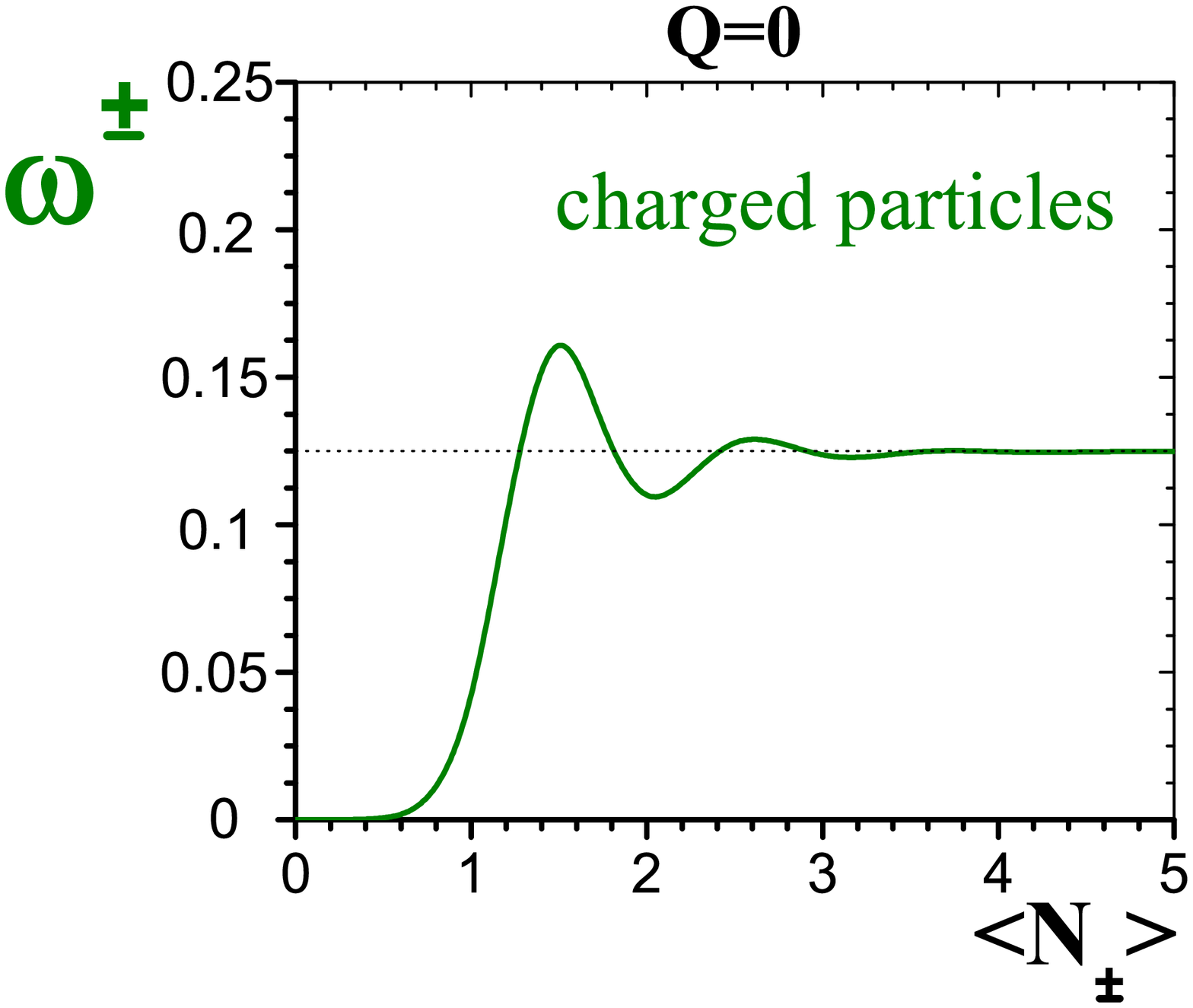}
 \vspace{-0.5cm}
 \caption{The scaled variances in the GMCE, left, and in MCE, right \cite{mce1}.}
 \label{fig-mce-gmce}
\end{center}
\end{figure}

The microcanonical partition function can be easily calculated
analytically for the system of $N$ noninteracting massless neutral
particles if one neglects the effects of quantum statistics. This
is just $N$-times integrated over momentum $\delta$-function
\cite{Fermi}:
%
\eq{\label{ZNgmce}
 Z_N(E,V) &\,=\, \frac{1}{N!}\left(\frac{gV}{2\pi^2}\right)^N
 \int_0^{\infty} p_1^2 dp_1\ldots \int_0^{\infty}p^2_N dp_N\;\;
 \delta\left(E-\sum^N_{j=1}p_j\right)
  \nonumber
\\
 &\,=\, \frac{1}{N!} \left(\frac{gV}{\pi^2}\right)^N
\frac{E^{3N-1}}{(3N - 1)!}
 }
where 
$E$ - is the energy and $V$ - volume of the system. One can also
generalize Eq.~(\ref{ZNgmce}) to the system of charged particles
\cite{mce1}:
 \eq{\label{ZNmce}
 Z_{N_+,N_-}(E,V,Q)
 \;=\; \frac{1}{N_+!N_-!}
 \left(\frac{gV}{\pi^2}\right)^{N_++N_-}\!\!\!
       \frac{E^{3(N_++N_-)-1}}{[3(N_++N_-) - 1]!}\;
       \delta(Q-[N_+-N_-])\;,
 }
and calculate corresponding scaled variances using
Eqs.~(\ref{ZNgmce}) and (\ref{ZNmce}) similarly to (\ref{<N>ce}),
(\ref{<N2>ce}). In thermodynamic limit, $V\rightarrow\infty$, and
for $Q=0$ it gives \cite{mce1}:
\eq{\label{W-mce-gmce}
 \omega_{g.m.c.e.} \simeq \frac{1}{4}
 \left(1\,-\,\frac{1}{8\,\langle N\rangle }\,+\, \dots \right),\qquad
 \omega_{m.c.e.}^{\pm}(Q=0) \simeq \frac{1}{8}
 \left(1\,-\,\frac{49}{1152\, \langle N_{\pm}^2\rangle }\,+\,... \right)\;,
 }
where $\langle N\rangle$ and $\langle N_\pm\rangle$ are the
average number of particles in GCE. Thus, one can see that the
scaled variance in thermodynamic limit is 4 and 8 times smaller
than in GCE for GMCE and MCE correspondingly, see
Fig.~\ref{fig-mce-gmce}.

It means that the thermodynamic equivalence for mean particle
number does not apply to fluctuations measured in terms of the
scaled variance \cite{ce1}-\cite{ce-Keranen}:
 \eq{
 \langle N\rangle \;\simeq\; \langle N\rangle_{c.e.} \;\simeq\; \langle
 N\rangle_{m.c.e.}&\;,\qquad V\rightarrow\infty
 \\
 \omega_{g.c.e.}  \;\neq\; \omega_{c.e.} \;\neq\; \omega_{m.c.e.}&\;,\qquad
 V\rightarrow\infty
 }
see also \cite{New-Trends} for the summary of some limiting values
of the scaled variance. Note, that average particle numbers in
GMCE, MCE and GCE are equivalent in thermodynamic limit
\cite{mce1} similarly to CE, see (\ref{TDlimit-CE}),
(\ref{TDlimit-CE2}), left. Canonical and microcanonical
suppression \cite{ce1}, \cite{mce1}, \cite{ce2} and even
microcanonical enhancement \cite{Threshold} of average
multiplicity $\langle N\rangle$ is observed for very small systems
only. Quantitatively, the limiting behavior in the MCE is reached
even quicker than in CE: for $2\div 3$ particles if we consider
$\langle N\rangle$ or $\langle N_{\pm}\rangle$ and for $3\div 4$
particles if we consider scaled variance see
Fig.~\ref{fig-mce-gmce} and \cite{mce1}.

The analytic calculations presented above are possible only for
Boltzmann statistic in CE and for Boltzmann massless particles in
MCE. The inclusion of other conserved charges and quantum
statistic makes the calculations technically very difficult. The
simplest way to overcome these difficulties is to consider
multiplicity distributions in different ensembles~\cite{CLT}.

\section{Multiplicity distribution}
Multiplicity distribution\footnote{probability to find some number
of particles $N$ if their average number $\langle N\rangle$ is
fixed by external conditions.}, partition function, different
moments, variance and scaled variance are closely related, namely:
 \eq{\label{P(N)}
 P(N) &\;\equiv\; \frac{Z_N}{Z}\;,&
 \langle N^k\rangle &\;\equiv\; \sum_N N^k P(N)\;,
 \\
 \langle (\Delta N)^2\rangle
   &\;\equiv\; \langle N^2\rangle \;-\; \langle N\rangle^2\;, &
 \omega
   &\;\equiv\; \frac{\langle (\Delta N)^2\rangle}{\langle N\rangle} \;.
 }
Multiplicity distribution $P(N)$ in ideal gas tends to Gaussian
$P_G(N)$ for $N\gg 1$:
 \eq{
 P(N\gg 1) \;\simeq\; P_G(N) = \frac{1}{\sqrt{2 \pi\,  \omega\cdot\langle N
 \rangle}} \,\exp \left[ - \frac{\left(N - \langle N \rangle
 \right)^2} {2 \, \omega \cdot\langle N \rangle} \right]~,
 }

\begin{figure}[h!]
\begin{center}
\includegraphics[width=0.6\textwidth]{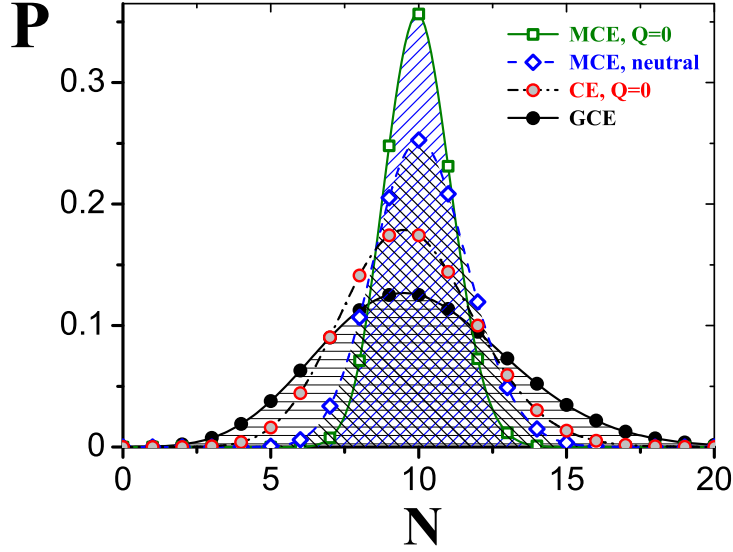}
 \vspace{-0.7cm}
\caption{Multiplicity distributions in MCE, GMCE, CE and GCE (from
top to bottom) calculated by means of (\ref{ZNmce}),
(\ref{ZNgmce}), (\ref{Zce}) and (\ref{Zgce})
correspondingly.}\label{fig-P(N)}
\end{center}
\end{figure}

One can easily check this for Eqs.~(\ref{Zgce}), (\ref{Zce}) and
(\ref{ZNgmce}), (\ref{ZNmce}), see the result in
Fig.~\ref{fig-P(N)} and detailed calculations in CE \cite{ce1},
MCE and GMCE \cite{mce1}.

One can see that multiplicity distributions in different ensembles
have the same maximum at $N=\langle N\rangle$, but different
width\footnote{in non-relativistic case $N=const$ by definition
and $P(N)\sim\delta(N)$ i.e. it would be a vertical line in
Fig.~\ref{fig-P(N)}.}. As an example in Fig.~\ref{fig-P(N)} we
choose $N=\langle N\rangle=10$. One can also see that the
distributions are smooth and have Gaussian form. Thus,
quantitatively, $N=10$ is already big enough to consider Gaussian
approximation.

To generalize our formalism for several conserved charges and
include quantum statistic, let us consider again a gas of
Boltzmann particles in the CE for simplicity. The GCE and CE
partition function are also closely related:
\eq{\label{Zce-gce}
 Z(T,V,\mu_Q)
 \;= \sum_{Q=-\infty}^{\infty} e^{Q\mu_Q/T} \sum_{N_+,N_-}Z_{N_+,N_-}(T,V,Q)
 }
The substitution of (\ref{Zce}) in (\ref{Zce-gce}) transforms it
to the identity \cite{Abramowitz}:
 \eq{
 Z(T,V,\mu_Q)
 &\;= \sum_{Q=-\infty}^{\infty} e^{Q\mu_Q/T} I_Q(2z)
  \;=\; \exp\left(2z\cosh[\mu_Q/T]\right)\;.
  \nonumber
}
After the replacement $e^{Q\mu_Q/T}=e^{(N_+-N_-)\mu_Q/T}$ and
$z_{\pm}=z\,e^{\pm\mu_Q/T}$ one obtains:
 \eq{
 Z(T,V,\mu_Q) \;\equiv\; Z
 &\;=\; \sum_{Q=-\infty}^{\infty}
        \sum_{N_+,N_-} \frac{z_+^{N_+}}{N_+!}\;
        \frac{z_-^{N_-}}{N_-!}
        \;\delta (Q-[N_+-N_-])
 \nonumber
 \\
 &\;=\; \sum_{Q=-\infty}^{\infty} \int_{-\pi}^{+\pi} \frac{d\phi}{2\pi} \;\;
        \exp\left[ -iQ\phi \;+\; z\left(e^{\mu_Q/T+i\phi}\;+\;e^{-\mu_Q/T-i\phi}\right)\right]
 \nonumber
 \\
 &\;=\; \sum_{Q=-\infty}^{\infty} \int_{-\pi}^{+\pi} \frac{d\phi}{2\pi}\;\;
        e^{-iQ\phi}\; \mathcal{Z}(\phi)
  \;=\; \sum_{Q=-\infty}^{\infty}  Z_Q\;,
}
where $\mathcal{Z}(\phi)$ is the GCE partition function with
replaced chemical potential $\mu_Q/T\rightarrow\mu_Q/T+i\phi$.
Similarly to (\ref{P(N)}), the probability of finding the GCE
system with the particular net-charge $Q$ equals the following
\cite{CLT}:
 \eq{\label{P(Q)}
 P(Q) \;=\; 
      \frac{1}{Z}\int_{-\pi}^{+\pi} \frac{d\phi}{2\pi}\;\;
        e^{-iQ\phi}\; \mathcal{Z}(\phi)
      \;=\; \frac{e^{Q\mu_Q/T}}{Z}\; I_Q(2z)\;.
 }
The probability to find the number of positively charged particles
$N_+$ that is exactly equal to $N$ in the GCE system with
net-charge equal to $Q$ is as follows~\cite{CLT}:
 \eq{\label{P(N,Q)}
 P(N,Q)
 &\;=\; \frac{1}{Z}\sum_{N_+,N_-} \frac{z_+^{N_+}}{N_+!}\;\frac{z_-^{N_-}}{N_-!}
         \;\delta (Q-[N_+-N_-])\;\delta (N-N_+])
  \nonumber
  \\
 &\;=\; \frac{1}{Z}\int_{-\pi}^{+\pi} \frac{d\phi}{2\pi}
       \int_{-\pi}^{+\pi} \frac{d\phi_N}{2\pi}\;\;
        e^{-iQ\phi}e^{-iN\phi_N}\; \mathcal{Z}(\phi,\phi_N)
  \;=\; \frac{e^{Q\,\mu_Q/T}}{Z}~\frac{z^{2N-Q}}{N!(N-Q)!}\; ,
 }
where $\mathcal{Z}(\phi,\phi_N)=\exp\left[
z\left(e^{\mu_Q/T+i\phi+i\phi_N}+e^{-\mu_Q/T-i\phi}\right)\right]$.
Finally, the particle number distribution in CE can be found as a
ratio of the distributions (\ref{P(N,Q)}) and (\ref{P(Q)})
calculated in GCE~\cite{CLT}:
 \eq{\label{P(N|Q)}
  P(N|Q)~ =~ \frac{P(N,Q)}{P(Q)}
 }
One can easily check that
 \eq{
 \langle N^k_+\rangle_{c.e.} \;=\; \sum_N N^k P(N|Q)
 \;=\; \frac{1}{I_Q\left(2z \right)}
        \sum_N N^k\; \frac{z^{N}}{N!} ~
        \frac{z^{N-Q}}{(N-Q)!}\;.
 }
The Eq.~(\ref{P(N|Q)}) is very important, because it allows to
calculate a value in CE using the values calculated in GCE. It
also allows for generalization to quantum statistic and taking
into account several exactly conserved charges, energy
conservation, resonance decay, etc. To do this one just need to
take corresponding GCE partition function and multiply it by the
Fourier representations of the relevant delta functions
\cite{CLT}:
 \eq{\label{P(Q^j)}
 P(Q^j) = \frac{1}{Z}\left[ \prod_{j} \int \limits_{-\pi}^{\pi}
  \frac{d\phi_j}{\left( 2\pi \right)} \right]~ e^{-iQ^j\phi_j}\;
  \mathcal{Z}(\phi_j)~,
 }
where $j$ runs over all conserved quantities. Repeated upper and
lower indexes $j$ imply summation over $j$. The function
$\mathcal{Z}(\phi)$ also changes if we include different particle
species and quantum statistic:
 \eq{
 \mathcal{Z}(\phi_j) \;=\; \exp \left[ \sum_{l} z_l \left( \phi_j \right) \right]~,
 }
where the single particle partition function of particle specie
$l$ is given by:
\begin{eqnarray}\label{QstatsZET}
z_l \left(\phi_j \right)
 &\;=\;&  \frac{g_l V} {\left( 2 \pi\right)^3}
           \int d^3p ~\ln\left[
            \left( 1 \pm  e^{-\left(\varepsilon_l -\mu_l \right) /T}\;
           \; e^{ i q^j_l \phi_j}\right)^{\pm1}\right]
 ~\equiv~ V \psi_l \left(\phi_j \right).
\end{eqnarray}
We introduced here particle $l$'s charges $q^j_l = \vec q_l =
(q_l,b_l,s_l,...)$ that corresponds to the charges conserved in
the system. We also introduced the degeneracy factor
$g_l=(2J_l+1)$, internal angular momentum $J_l$, mass $m_l$, and
energy $\varepsilon_l=\sqrt{p^2+m_l^2}$, the chemical potential
vector ${\mu^j} = (\mu_Q,\mu_B,\mu_S...)$, and particle $l$'s
chemical potential $\mu_l = q^j_l \mu_j $. $V$ is the system
volume, and $T$ it's temperature. The summation $\sum_l$ includes
also anti-particles, for which $q^j_l\rightarrow -q^j_l$. The
upper sign in the Eq.~(\ref{QstatsZET}) denotes Fermi-Dirac
statistics, while the lower is used for Bose-Einstein statistics.
The Boltzmann approximation is obtained from (\ref{QstatsZET}) as
a first term of the series expansion for $e^{-\left(\varepsilon_l
-\mu_l \right) /T}\ll 1$.

The real calculations of (\ref{P(Q^j)}) can be performed only in
the limit $V\rightarrow\infty$. Then the main contribution to the
integral in (\ref{P(Q^j)}) comes from a small region around the
origin. Thus it is possible to make the Taylor expansion of
$\sum_l\psi_l$ and leave only the first two terms. Similar saddle
point expansion was intensively used for partition function itself
\cite{CE-Res}, \cite{ce-Keranen}, \cite{saddle_point},
\cite{saddle_point0} while the relations between partition
function, multiplicity distribution, and scaled variance was
obtained only in \cite{CLT}.

It was shown for GMCE in \cite{mce-log} and for the most general
case of MCE with arbitrary number of conserved charges in
\cite{CLT} that the variance is proportional to the ratio of
correlation matrix determinants:
 \eq{
 \langle (\Delta N)^2\rangle \;=\; V\,\frac{\det|\,\widetilde{A}\,|}{\det|\,A\,|}
 }
where the elements of the correlation matrixes can be found as
follows:
 \eq{\label{Aij}
 A_{i,j}
   \;=\; -\frac{\partial^2\log \mathcal{Z}(\phi_j)}
               {\partial \phi_i\, \partial\phi_j}
         \bigg|_{\vec{\phi}=0}\;\;, &&
 \widetilde{A}_{i,j}
   \;=\; \frac{\partial^2\log \mathcal{Z}(\phi_j,\phi_N)}
              {\partial \phi_i\, \partial\phi_j}
         \bigg|_{\vec{\phi},\phi_N=0}.
 }
The difference between $A$ and $\widetilde{A}$ is that in the
latter case $i$ and $j$ run over $N$ also \cite{CLT}. Then the
scaled variance is a ratio of (\ref{Aij}) to the mean
multiplicity:
 \eq{
 \langle N\rangle \;=\; -i\,
   \frac{ \partial\log \mathcal{Z}(\phi_j,\phi_N) }{\partial\phi_N}
         \bigg|_{\vec{\phi},\phi_N=0}\;.
 }

The above method is very powerful. Nevertheless it fails in the
case of Bose condensation, because scaled variance in GCE then
goes to infinity \cite{ce-BF} and multiplicity distribution has
infinite width. All matrix elements (\ref{Aij}) and higher
derivatives of $\mathcal{Z}(\phi)$ tends to infinity in GCE
\cite{CLT}. However, exact charge and energy conservation suppress
even these infinite fluctuations \cite{ce-BF}, \cite{BEC}. The
very special selection of events is need to see these infinite
fluctuations in MCE. This is proposed as the signal of possible
$\pi$-meson condensation in $p+p$ collisions \cite{BEC},
\cite{BEC-CPOD-2007}.

The further improvement is possible if one consider average
multiplicity and fluctuations at different momentum levels. This
approach is called the microcorrelator method \cite{mce1}. It
analogous to the above approach \cite{CLT}, but additionally
allows to consider correlations between different momentum levels.
The full hadron gas in the next section is considered using
microcorrelator method \cite{CE-Res}, \cite{CE-HSD}.


\section{Hadron Gas}

Let us consider the fluctuations in the ideal relativistic gas
with different types of hadrons in the MCE with exactly fixed the
global electric (Q), baryon (B), and strange (S) charges of the
statistical system.
The system of non-interacting Bose or Fermi particles of species
$i$ can be characterized by the occupation numbers $n_{p,i}$ of
single quantum states labelled by momenta $p$. The occupation
numbers run over $n_{p,i} = 0,\, 1$ for fermions and $n_{p,i} =
0,\, 1, 2, \ldots$ for bosons. The GCE average values and
fluctuations of $n_{p,i}$ equal the following \cite{Landau2}:
 \eq{
 \langle n_{p,i} \rangle
 ~& = ~\frac {1} {\exp \left[\left(\sqrt{p^{2}+m_i^{2}}~-~ \mu_i\right) / T\right]
 ~-~ \gamma_i}~, \label{np-aver}
 \\
 \upsilon^{ 2}_{p,i}
 ~& \equiv~ \langle\Delta n_{p,i}^2\rangle
 ~ \equiv~ \langle \left( n_{p,i}-\langle
 n_{p,i}\rangle\right)^2\rangle
 ~=~ \langle n_{p,i}\rangle
\left(1 + \gamma_i \langle n_{p,i} \rangle\right)~.\label{np-fluc}
  }
In Eq.~(\ref{np-aver}), $T$ is the system temperature, $m_i$ is
the mass of $i$-th particle species, $\gamma_i$ corresponds to
different statistics ($+1$ and $-1$ for Bose and Fermi,
respectively, and $\gamma_i=0$ gives the Boltzmann approximation),
and chemical potential $\mu_i$ equals:
\eq{ \mu_i~=~q_i~\mu_Q~+~b_i~\mu_B~+~s_i~\mu_S ~,\label{chempot}}
where $q_i,~b_i,~s_i$ are the electric charge, baryon number and
strangeness of particle of specie $i$, respectively, while
$\mu_Q,~\mu_B,~\mu_S$ are the corresponding chemical potentials
which regulate the average values of these global conserved
charges in the GCE.

The average number of particles of species $i$, the number of
positively and negatively charged particles are equal:
 \eq{\label{Ni-gce}
 \langle N_i\rangle \;& \equiv\; \sum_p \langle n_{p,i}\rangle
 \;=\; \frac{g_i V}{2\pi^{2}}\int_{0}^{\infty}p^{2}dp\; \langle
 n_{p,i}\rangle\;,
 &&
  \langle N_{+}\rangle \;=\; \sum_{i,q_i>0} \langle N_i\rangle\;,
 &&
 \langle N_-\rangle \;=\; \sum_{i,q_i<0} \langle N_i\rangle\;,
 %
 }
where $g_i$ is the degeneracy factor of particle of species $i$. A
sum of the momentum states means 
the momentum integral, which holds in the thermodynamic limit
$V\rightarrow \infty$.

Particle number fluctuations and correlations can be calculated in
all ensembles using the microscopic correlator method.
 \eq{
  \langle \Delta N_i ~\Delta N_j~\rangle_{\ldots}
 &~= \sum_{p,k}~\langle \Delta n_{p,i}~\Delta
 n_{k,j}\rangle_{\ldots}\;, \label{mc-corr-mce}
 }
where $\langle\;\;\; \rangle_{\ldots}$ means GCE, CE, or MCE
microscopic correlator.
The scaled variances of
negatively and positively charged particles read:
\eq{\label{omega-all}
 \omega^- ~=~ \frac{\langle \left( \Delta N_- \right)^2
\rangle}{\langle N_-
  \rangle}~,
~~~~\omega^+~ =~ \frac{\langle \left( \Delta N_+ \right)^2
\rangle}{\langle N_+
  \rangle}~,
}
where
\eq{
\langle \left( \Delta N_- \right)^2 \rangle~ =
\sum_{i,j;~q_i<0,q_j<0} \langle \Delta N_i \Delta N_j \rangle~,& &
\langle \left( \Delta N_+ \right)^2 \rangle~ =~
\sum_{i,j;~q_i>0,q_j>0} \langle \Delta N_i \Delta N_j \rangle~.
\label{DNpm} }
The microscopic correlator in the GCE reads:
\eq{\label{corr-gce}
 \langle \Delta n_{p,i} ~ \Delta n_{k,j}
\rangle
 \;=\;  \upsilon_{p,i}^2\,\delta_{ij}\,\delta_{pk}~,
 }
where $\upsilon_{p,i}^2$ is given by Eq.~(\ref{np-fluc}). This
gives a possibility to calculate the fluctuations of different
observables in the GCE. Note that only particles of the same
species, $i=j$, and on the same level, $p=k$, do correlate in the
GCE. Thus, Eq.~(\ref{corr-gce}) is equivalent to
Eq.~(\ref{np-fluc}): only the Bose and Fermi effects for the
fluctuations of identical particles on the same level are relevant
in the GCE.

The MCE microscopic correlator is as follows \cite{CE-Res},
\cite{CE-HSD}:
 \eq{\label{corr}
 &
 \langle \Delta n_{p,i}  \Delta n_{k,j} \rangle_{m.c.e.}
 ~=\;  \upsilon_{p,i}^2\,\delta_{ij}\,\delta_{pk}
 \;-\;  \frac{\upsilon_{p,i}^2\upsilon_{k,j}^2}{|A|}\;
 [\;q_iq_j M_{qq} + b_ib_j M_{bb} + s_is_j M_{ss}
 \\
 &+ ~\left(q_is_j + q_js_i\right) M_{qs}~
 - ~\left(q_ib_j + q_jb_i\right) M_{qb}~
 - ~\left(b_is_j + b_js_i\right) M_{bs}\nonumber
 \\
 &+~ \epsilon_{pi}\epsilon_{kj} M_{\epsilon\epsilon}~-~ \left(q_i \epsilon_{pj} + q_j\epsilon_{ki} \right)
 M_{q\epsilon}~
  +~ \left(b_i \epsilon_{pj} + b_j\epsilon_{ki} \right)
  M_{b\epsilon}~
  - ~\left(s_i \epsilon_{pj} + s_j\epsilon_{ki} \right) M_{s\epsilon}
 \;]\;, \nonumber
 }
where $|A|$ is the determinant and $M_{ij}$ are the minors of the
following matrix:
 \eq{\label{matrix}
 A =
 \begin{pmatrix}
 \Delta (q^2) & \Delta (bq) & \Delta (sq) & \Delta (\epsilon q)\\
 \Delta (q b) & \Delta (b^2) & \Delta (sb) & \Delta (\epsilon b)\\
 \Delta (q s) & \Delta (b s) & \Delta (s^2) & \Delta (\epsilon s)\\
 \Delta (q \epsilon) & \Delta (b \epsilon) & \Delta (s \epsilon) & \Delta (\epsilon^2)
 \end{pmatrix}\;,
 }
with the elements, $\;\Delta (q^2)\equiv\sum_{p,k}
q_k^2\upsilon_{p,k}^2\;$, $\;\Delta (qb)\equiv \sum_{p,k}
q_kb_k\upsilon_{p,k}^2\;$, $\;\Delta (q\epsilon)\equiv \sum_{p,k}
q_k\epsilon_{pk}\upsilon_{p,k}^2\;$, etc. The sum, $\sum_{p,k}$~,
means integration over momentum $p$, and summation over all
hadron-resonance species~$k$ contained in the model. Note that the
presence of MCE terms containing single particle energies,
$\epsilon_{pi}=\sqrt{p^{2}+m_j^{2}}$, in the last line of
Eq.(\ref{corr}) is a consequence of exact energy conservation. In
the CE, only charges are conserved exactly, thus the terms of the
last line in Eq.~(\ref{corr}) are absent, and  $A$ in
Eq.~(\ref{matrix}) becomes the $3\times 3$ matrix (see
Ref.~\cite{CE-Res}).
%
%

\section{Effect of resonance decays}
The average number of $i$-particles in the presence of primary
particles $N_i^*$ and different resonance types $R$ is the
following:
 \eq{\label{<N>}
 \langle N_i\rangle
 \;=\; \langle N_i^*\rangle + \sum_R \langle N_R\rangle \sum_r b_r^R n_{i,r}^R
 \;\equiv\; \langle N_i^*\rangle + \sum_R \langle N_R\rangle \langle n_{i}\rangle_R
 }
The summation $\sum_R$ runs over all types of resonances. The
$\langle\ldots\rangle$ and $\langle\ldots\rangle_R$ correspond to
the GCE averaging, and that over resonance decay channels.
Resonance decay has a probabilistic character. This itself causes
the particle number fluctuations in the final state. In the GCE
the final state correlators can be calculated as \cite{Koch}:
 \eq{\label{corr-GCE}
  \langle \Delta N_i\,\Delta N_j\rangle
  ~=~
  \langle\Delta N_i^* \Delta N_j^*\rangle
  \;+\; \sum_R \left[ \langle\Delta N_R^2\rangle\;
  \langle n_{i}\rangle_R\;\langle n_{j}\rangle_R
  \;+\; \langle N_R\rangle\; \langle \Delta n_{i}\Delta n_{j}\rangle_R
  \right]~,
  }
where $b^R_r$ is the branching ratio of the $r$-th branch,
$n_{i,r}^R$ is the number of $i$-th particles produced in that
decay mode, and $r$ runs over all branches  with the requirement
$\sum_{r} b_r^R=1$ and $\langle \Delta n_i~\Delta
n_j\rangle_R\equiv \sum_r b_r^R n_{i,r}^R n_{j,r}^R~-~\langle
n_i\rangle_R\langle n_j\rangle_R$~. Note that different branches
are defined in a way that final states with only stable (with
respect to strong and electromagnetic decays) hadrons are counted.

All primary particles and resonances become to correlate in the
presence of exact charge conservation laws. Thus for the MCE
correlators
we obtain a new result \cite{CE-Res}:
 \eq{\label{corr-CE}
 & \langle \Delta N_i\,\Delta N_j\rangle_{m.c.e.}
 \;=\; \langle\Delta N_i^* \Delta N_j^*\rangle_{m.c.e.}
  \;+\; \sum_R \langle N_R\rangle\; \langle \Delta n_{i}\; \Delta  n_{j}\rangle_R
 \;+\; \sum_R \langle\Delta N_i^*\; \Delta N_R\rangle_{m.c.e.}\; \langle n_{j}\rangle_R
  \; \nonumber
 \\
 &+\; \sum_R  \langle\Delta N_j^*\;\Delta N_R\rangle_{m.c.e.}\; \langle n_{i}\rangle_R
  \;+\; \sum_{R, R'} \langle\Delta N_R\;\Delta N_{R'}\rangle_{m.c.e.}
  \; \langle n_{i}\rangle_R\;
       \langle n_{j}\rangle_{R^{'}}\;.
 }
Additional terms in Eq.~(\ref{corr-CE}) compared to
Eq.~(\ref{corr-GCE}) are due to the correlations induced by exact
charge conservations in the MCE. The Eq.~(\ref{corr-CE}) remains
valid in the CE too with $\langle \ldots \rangle_{m.c.e.}$
replaced by $\langle \ldots \rangle_{c.e.}$, the difference
between them appears only when one specifies the microscopic
correlators (\ref{corr}) of the MCE or CE.

\section{Scaled variances along the chemical freeze-out line}
Mean hadron multiplicities in heavy ion collisions at high
energies can be approximately fitted by the GCE hadron-resonance
gas model. The fit parameters are temperature $T$, chemical
potentials ($\mu_B$, $\mu_S$, $\mu_Q$), and strangeness
suppression factor $\gamma_S$, which allows for non-equilibrium
strange hadron yields. There are several programs designed for the
analysis of particle multiplicities in relativistic heavy-ion
collisions within the hadron-resonance gas model, see e.g., SHARE
\cite{Share}, THERMUS \cite{Thermus} and THERMINATOR
\cite{Therminator}. In this paper an extended version of the
THERMUS thermal model framework \cite{Thermus} is used.

\begin{figure}
\begin{center}
\includegraphics[width=0.6\textwidth]{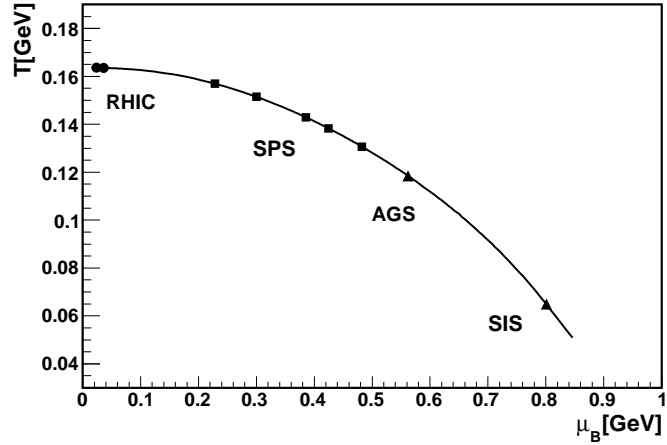}
 \vspace{-0.5cm}
 \caption{The chemical freeze-out line in central
A+A collisions \cite{CE-Res}.}\label{fig_T_muB}
\end{center}
\end{figure}

For the chemical freeze-out condition we choose the average energy
per particle $\langle E\rangle /\langle N\rangle = 1 GeV$
\cite{Cl-Red}. Using the standard parametrization \cite{param} we
obtain the $T-\mu_B$ freeze-out line for central A+A collisions
(see Fig.~\ref{fig_T_muB}).
The center of mass nucleon-nucleon energies, $\sqrt{S_{NN}}$,
marked in the figures below correspond to the beam energies at SIS
(2A GeV), AGS (11.6A GeV), SPS (20A, 30A, 40A, 80A, and 158A GeV),
colliding energies at RHIC ($\sqrt{S_{NN}}$ = 62.4 GeV, 130 GeV
and 200 GeV) and LHC ($\sqrt{S_{NN}}$ = 5500 GeV).
\begin{figure}[ht!]
\begin{center}
\includegraphics[width=0.49\textwidth]{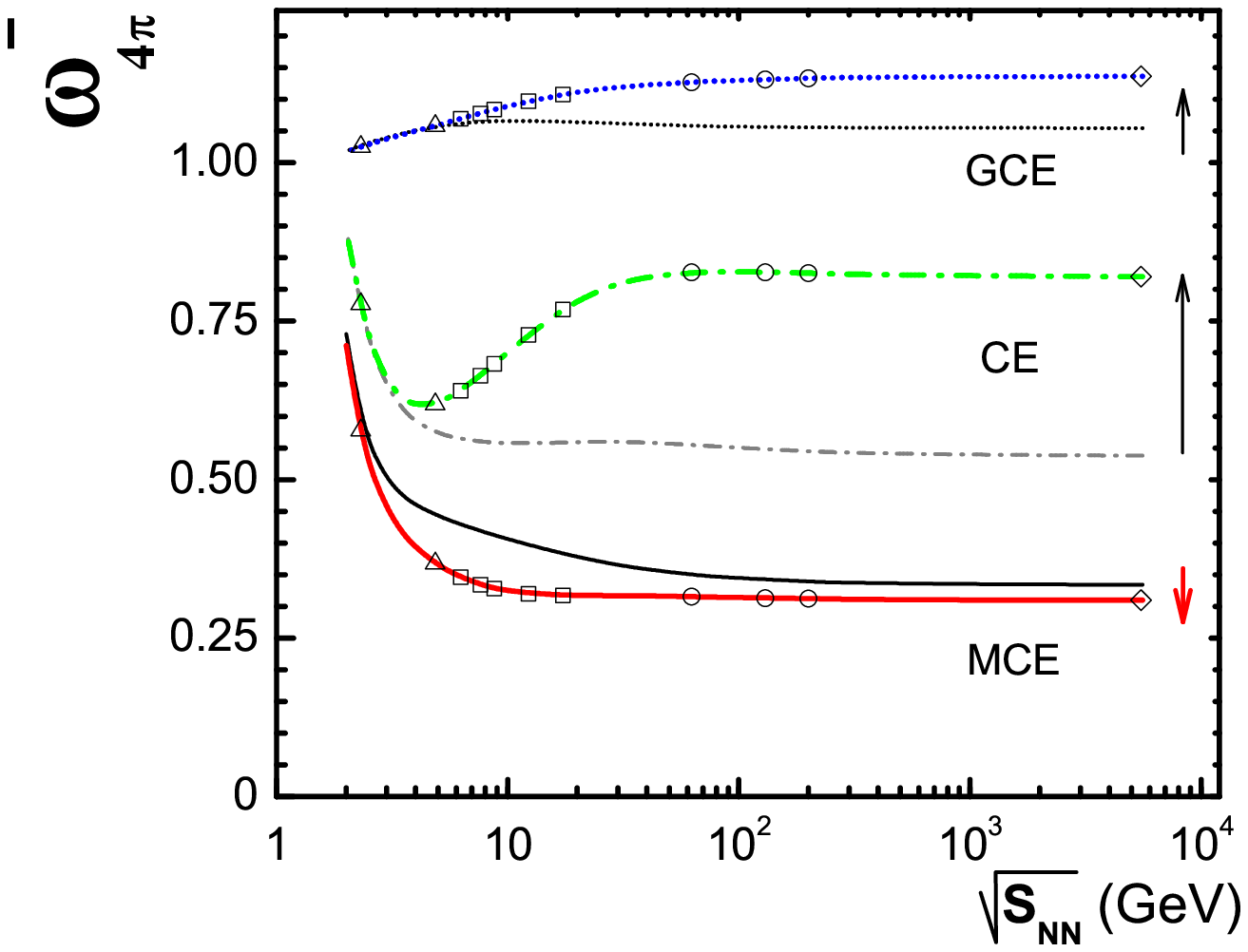}
\includegraphics[width=0.49\textwidth]{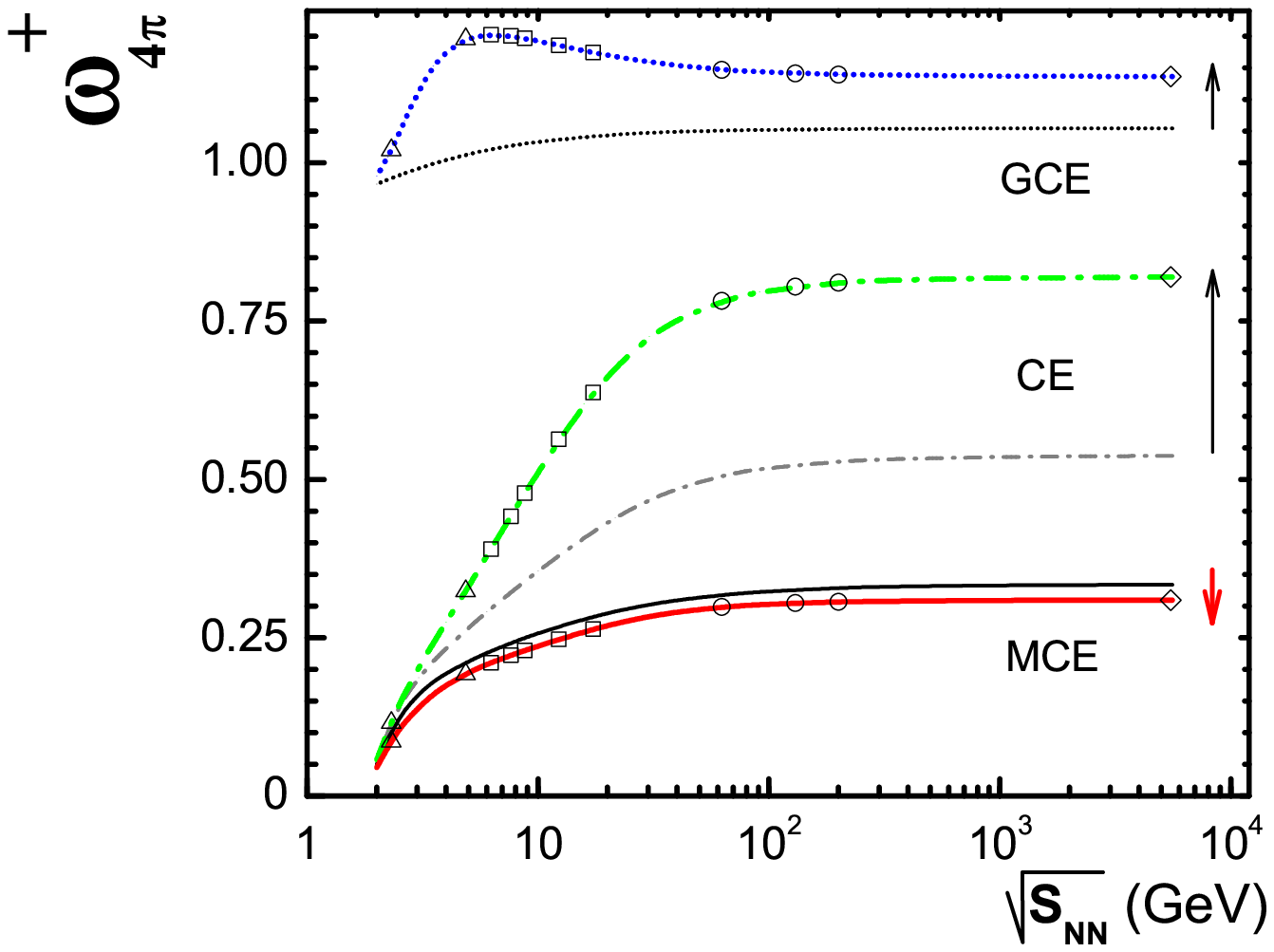}
 \vspace{-0.5cm}
 \caption{The scaled variances for negatively and positively charged
particles, both primordial and final, along the chemical
freeze-out line for central Pb+Pb (Au+Au) collisions. Different
lines present the GCE, CE, and MCE results. Symbols at the CE and
MCE lines for the final particles correspond to the specific
collision energies. The arrows show the effect of resonance decays
\cite{CE-HSD}.} \label{omega_m}
\end{center}
\end{figure}

Figure~\ref{omega_m} show the prediction for the scaled variances
for negatively and positively charged particles as a function of
$\sqrt{s_{NN}}$.

The prediction can be compared with the preliminary NA49 data on
Pb+Pb collisions at 20A-158A GeV \cite{Lungwitz} using the
following approximate formula:
 \eq{
 \omega^{\pm}_{acc} = 1 - q + q\, \omega^{\pm}_{4\pi} ,
  }
where $\omega_{4 \pi}$ refers to an ideal detector with full
$4\pi$-acceptance and $\omega^{\pm}_{acc}$ is the scaled variance
measured by a real detector with a limited acceptance), q is the
ratio between mean multiplicities of accepted particles and all
hadrons. In the limit of a very `bad' (or `small') detector,
$q\rightarrow 0$, all scaled variances approach linearly to 1,
i.e., this would lead to the Piossonian distributions for detected
particles. However, we find a strong qualitative difference
between the predictions of the statistical model valid for any
freeze-out conditions and experimental acceptances: the CE and MCE
correspond to $\omega^{\pm}_{m.c.e.} < \omega^{\pm}_{c.e.} < 1$,
and the GCE to $\omega^{\pm}_{g.c.e.}>1$.

\begin{figure}[ht!]
\begin{center}
\includegraphics[width=0.49\textwidth]{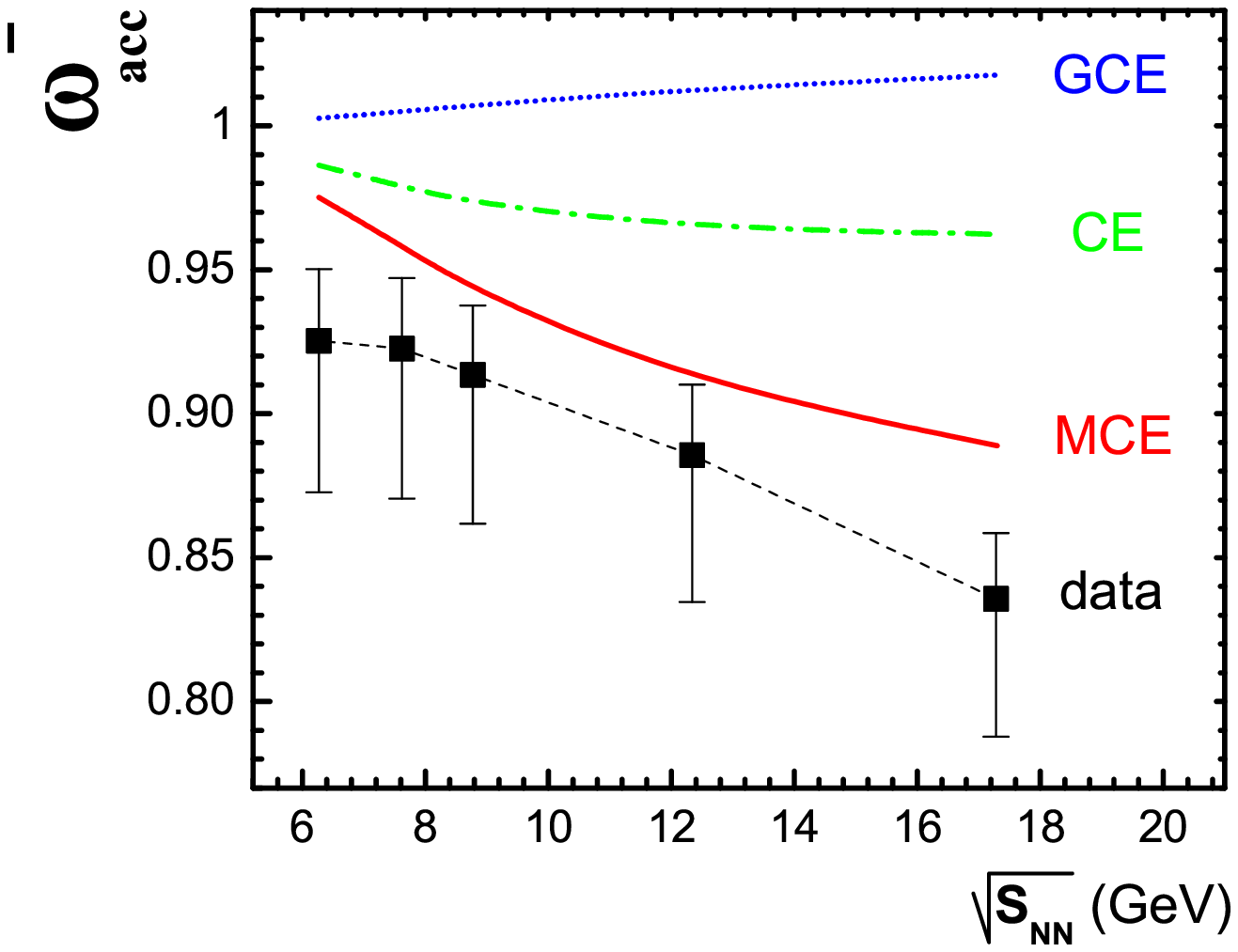}
\includegraphics[width=0.49\textwidth]{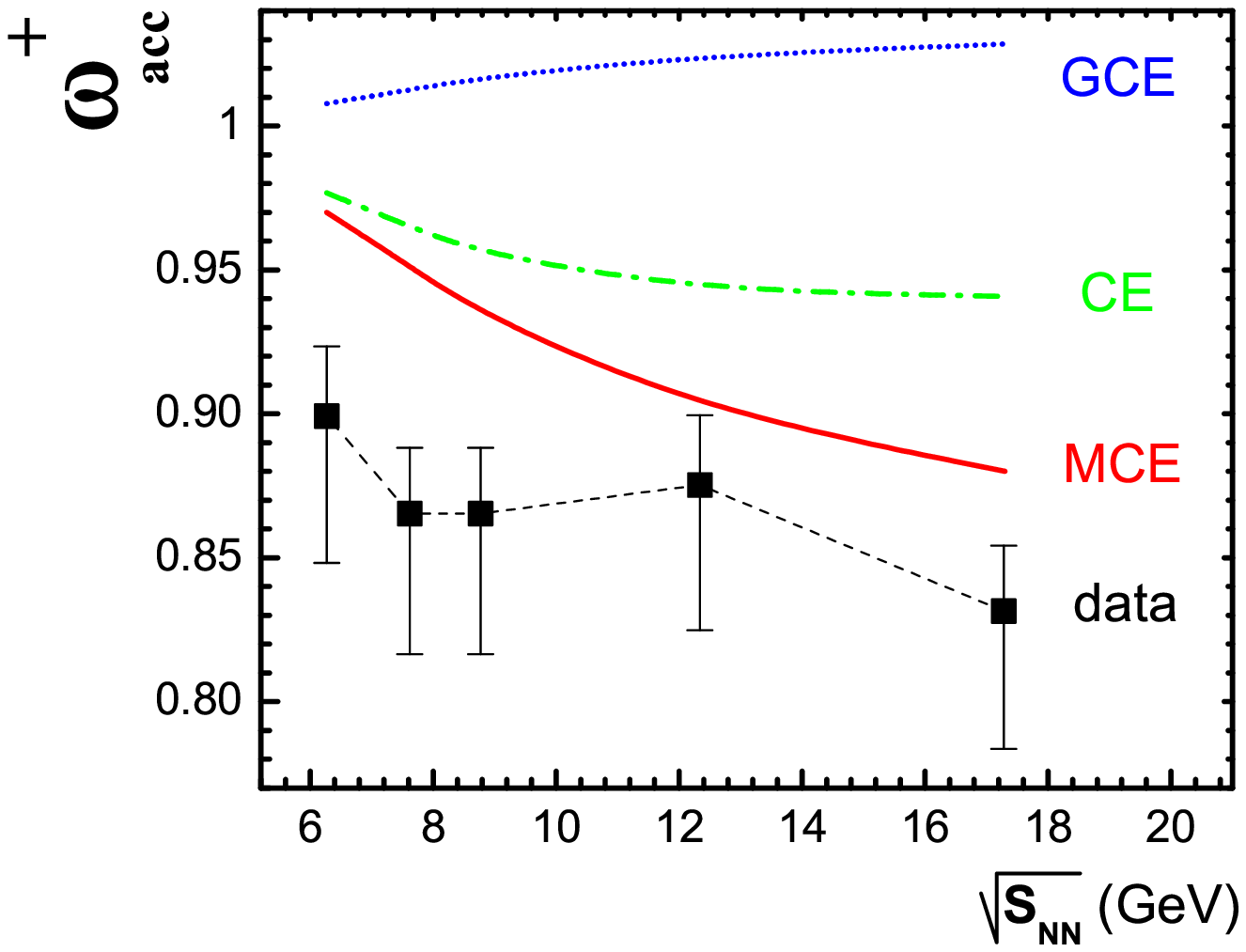}
 \vspace{-0.5cm}
\caption{The scaled variances for negative (left) and positive
(right) hadrons along the chemical freeze-out line for central
Pb+Pb collisions at the SPS energies. The corresponding $T$ and
$\mu_B$ values at different SPS collision energies are presented
in Fig.~\ref{fig_T_muB}. Different lines show the GCE, CE, and MCE
results calculated with the NA49 experimental acceptance
\cite{CE-HSD}.}\label{fig_omega_exp}
\end{center}
\end{figure}

From Fig.~\ref{fig_omega_exp} it follows that the NA49 data for
$\omega^{\pm}$ extracted from the most central Pb+Pb collisions at
all SPS energies are close to the results of the hadron-resonance
gas statistical model within the MCE. The data reveal even
stronger suppression of the particle number fluctuations. A
possible reason of this is an uncertainty in the determination of
the detector acceptance and  an additional suppression due to
momentum conservation and the excluded volume effects in the
hadron-resonance gas.

\begin{figure}[h!]
\begin{center}
\includegraphics[width=1\textwidth]{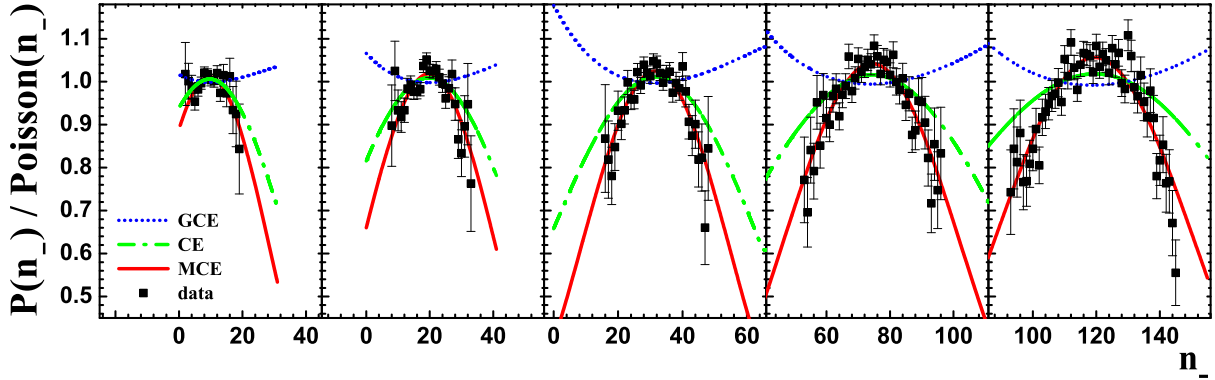}
 \vspace{-1.3cm}
\caption{The ratio of the multiplicity distributions to Poisson
ones for negatively charged hadrons produced in central (1$\%$)
Pb+Pb collisions at 20A GeV, 30A GeV, 40A GeV, 80A GeV, and 158A
GeV (from left to right) in the NA49 acceptance \cite{Lungwitz}.
The preliminary experimental data (solid points) of NA49
\cite{Lungwitz} are compared with the prediction of the
hadron-resonance gas model obtained within different statistical
ensembles, the GCE (dotted lines), the CE (dashed-dotted lines),
and the MCE (solid lines) \cite{CE-HSD}.}\label{fig-P-NA49}
\end{center}
\end{figure}

In order to allow for a detailed comparison of the distributions
the ratio of the data and the model distributions to the Poisson
one is presented in Fig.~\ref{fig-P-NA49}.
The convex shape of the data reflects the fact that the measured
distribution is significantly narrower than the Poisson one. This
suppression of fluctuations is observed at all five SPS energies
and it is consistent with the results for the scaled variance
shown and discussed previously. The GCE hadron-resonance gas
results are broader than the corresponding Poisson distribution.
The ratio has a concave shape. An introduction of the quantum
number conservation laws (the CE results) leads to the convex
shape and significantly improves agreement with the data. Further
improvement of the agreement is obtained by the additional
introduction of the energy conservation law (the MCE results). The
measured spectra surprisingly well agree with the MCE predictions
\cite{CE-HSD}.

\section{Summary}

We have found that scaled variances are different in different
statistical ensembles. For relativistic one component Boltzmann
gas with zero charge in thermodynamic limit we analytically
obtained rather interesting limiting values: $\omega_{g.c.e.}=1,\;
\omega_{c.e.}=1/2,\; \omega_{g.m.c.e.}(m=0) = 1/4$ and
$\omega_{m.c.e.}(m=0)=1/8$. We also found an analytical method to
account for resonance decays. The formalism that allows to
consider any number of conserved charges and also energy
conservation in full hadron-resonance gas was developed.

The experimental data allows to exclude GCE for scaled variance.
They show reasonable agreement with CE and surprisingly well agree
with the expectations for the MCE. Thus the predicted suppression
of the multiplicity fluctuations in relativistic gases in the
thermodynamic limit due to conservation laws do exist.

\begin{acknowledgments}
I would like to thank my PhD supervisor Mark I. Gorenstein and
also my co-authors: F.~Becattini, L. Ferroni, M. Gazdzicki, M.
Hauer, A. Keranen, V. P. Konchakovski, A. P. Kostyuk and O. S.
Zozulya. I would like also to thank for the support The
International Association for the Promotion of Cooperation with
Scientists from the New Independent states of the Former Soviet
Union (INTAS), Ref. Nr. 06-1000014-6454.

\end{acknowledgments}


\end{document}